\def\year{2022}\relax
\documentclass[letterpaper]{article} 
\usepackage{aaai22}  
\usepackage{times}  
\usepackage{helvet}  
\usepackage{courier}  
\usepackage[hyphens]{url}  
\usepackage{graphicx} 
\usepackage{natbib}  
\usepackage{caption} 
\DeclareCaptionStyle{ruled}{labelfont=normalfont,labelsep=colon,strut=off} 
\usepackage{algorithm}
\usepackage{algorithmic}

\usepackage{newfloat}
\usepackage{listings}
\floatstyle{ruled}
\newfloat{listing}{tb}{lst}{}
\floatname{listing}{Listing}
\title{Incremental Knowledge Tracing from Multiple Schools}

\author {
    Sujanya Suresh,\textsuperscript{\rm 1}
    Savitha Ramasamy, \textsuperscript{\rm 2}
    P.N. Suganthan, \textsuperscript{\rm1}
    Cheryl Sze Yin Wong \textsuperscript{\rm 2}
}
\affiliations {
    \textsuperscript{\rm 1} School of Electrical and Electronic Engineering, Nanyang Technological University\\ 50 Nanyang Ave, Singapore 639798 \\
    \textsuperscript{\rm 2} Institute for Infocomm Research (I2R), A-STAR \\1 Fusionopolis Way, Singapore 138632\\
    SUJANYA001@e.ntu.edu.sg, ramasamysa@i2r.a-star.edu.sg, EPNSugan@ntu.edu.sg, Cheryl\_Wong@i2r.a-star.edu.sg
}

\usepackage{bibentry}

\begin{document}

\maketitle

\begin{abstract}
Knowledge tracing is the task of predicting a learner's future performance based on the history of the learner's performance. Current knowledge tracing models are built based on an extensive set of data that are collected from multiple schools. However, it is impossible to pool learner's data from all schools, due to data privacy and PDPA policies. 
Hence, this paper explores the feasibility of building knowledge tracing models while preserving the privacy of learners' data within their respective schools. This study is conducted using part of the ASSISTment 2009 dataset, with data from multiple schools being treated as separate tasks in a continual learning framework. The results show that learning sequentially with the Self Attentive Knowledge Tracing (SAKT) algorithm is able to achieve considerably similar performance to that of pooling all the data together. 
\end{abstract}

\section{Introduction}
\label{sec:intro}
\noindent The rising effect of the COVID-19 pandemic has imposed the need to distance socially. In a measure to reduce physical interactions, schools and educational institutes have shifted to online modes of teaching. This shift towards the use of online platforms has allowed easier storage of information on students' learning journey. These information can include student learning activities, performance and attentiveness in a virtual environment. The topic of Knowledge Tracing (KT), which predicts learner's future performance has gained considerable attention \cite{riid4}. The main objective of KT in online modes of teaching is to model student's mastery of skills and concepts based on their history of learning activities, and to track their performances in various tasks and exercises \cite{inproceedings}.

Data acquired from multiple schools, when made available for all, would allow one to build a more generalizable model. However, sharing of data from multiple schools is restricted due to data privacy.Hence, we explore the feasibility of learning continually from  data of students in multiple institutions, without the need to share data. To this end, we define the data from individual school as a task, and learn a sequence of task continually. 
\\

We demonstrate the proposed 
continual learning framework using the ASSISTment 2009 dataset, adopting the Self Attentive Knowledge Tracing (SAKT) model. Data from three different schools with similar number of learners were chosen for this study. The performance on the various tasks is found to differ based on the order of tasks, given that the tasks have varying characteristics. Based on the observations, the model would be able to generalise better if the the data from the first task has high variability. 

\section{Related Works}
\label{sec:related}

Knowledge Tracing can be viewed as a supervised learning task which gets the input of the exercise interactions {X= $x_1$,$x_2$,..,$x_t$}, to predict the future interactions $x_{t+1}$. The exercises attempted by a student at time t and the correctness of it is represented as an interaction, $x_t$ = ($e_t,r_t$). KT aims to predict the probability of the student giving a correct response to the next exercise, i.e. P($r_{t+1}$ = $1|e_{t+1}$, \textbf{X}). 
Deep learning strategies were the main focus for constructing KT models, like the Deep Knowledge Tracing (DKT) \cite{piech2015deep} and Recurrent Neural Networks (RNN) \cite{Sherstinsky_2020} . Dynamic key-value Memory Networks (DKVNM)\cite{zhang2017dynamic} have gained attention mainly because of their outperforming interpretability in data compared with the traditional methods. The success of transformer based models\cite{vaswani2017attention} resulted in state-of-the-art performances on using key-value matrices\cite{Shin_2021} to model the relationship of students' knowledge level and provided exercises.

\section{Incremental Knowledge Tracing From Multiple Schools}
\label{sec:iktms}

In this section, we first introduce the ASSISTment2009 dataset and the portion of the data that we use
for our experiments. This is followed by a brief description of the SAKT algorithm. Next, we explain the experimental setup in the continual learning framework. Finally, we provide the model details used in our experiments.  

\subsubsection{Dataset: ASSISTment2009}\label{assist_data}
\label{ssec:dataset}

\noindent ASSISTment 2009 \cite{C2} is a mastery learning skill builder dataset provided by the online tutoring platform, ASSISTment, which is widely used for knowledge tracing tasks. A student completes the assignment and masters a skill if specific criteria like answering three questions correctly are met. The students used for our model are characterised based on the historical data of their school\_id's, and the drift in data is visualised with the clustering of the problems attempted by users from each school. The ASSISTment dataset has plethora of features including the tutor mode, problem\_set type, hint\_count which are used in various applications like Clustering, Personalization of student parameters and Intelligent Tutoring system\cite{BKTLSTM} that involves the Wheel-Spinning problem, where the student gets struck in a situation of finding it difficult to learn a skill from a given problem set. 
In our problem, we focus on the primary parameters of school\_id and problem\_id. 

\begin{table}[h]
    \centering
    \begin{tabular}{|c|c|c|c|}
        \hline
         School ID & No. of & No. of unique & No. of   \\
         &  Learners & Questions & Responses \\
         \hline
         1998 & 95 & 3065 & 5617 \\
         5117 & 92 & 2728 & 9746 \\
         5049 & 94 & 7975 & 19106 \\
         \hline
    \end{tabular}
    \caption{Characteristics of the ASSISTment 2009 dataset in multiple schools}
    \label{tab:data_char}
\end{table}

As it can be observed from Table \ref{tab:data_char}, we consider 3 schools that are similar in the number of users and their responses, in the study. We undertake a task based continual learning approach.

\begin{figure}[htb]
  \centering
  \centerline{\includegraphics[width=7.0cm]{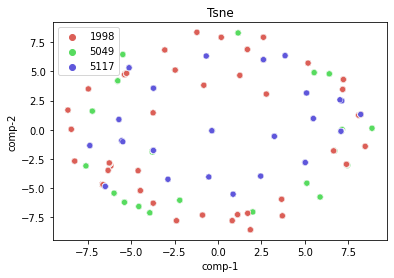}}
  \caption{t-SNE plot of the 3 tasks 5049,5117 and 1998 }\medskip
  \label{tsne}
\end{figure}

The t-distributed Stochastic Neighbour Embedding (tSNE) is a non linear dimensionality reduction method that helps separate data which cannot be done by a straight line. It rather projects the features onto a smaller space (2D in our case) to explore the data and its specifications. The t-SNE plot for the tasks 5049, 5117 and 1998 is depicted in Fig\ref{tsne}. The following observations are made about the data set, from the Fig. \ref{tsne}.
\begin{itemize}
\item 5049 and 1998 are more similar in terms of data distribution of problem\_id answered by the students
\item The data distribution of 5117 is different from 5049 and 1998
\end{itemize}

\subsection{Algorithm: Self Attention Knowledge Tracing}
\label{ssec:sakt}
SAKT is the first knowledge tracing model that utilizes transformer’s self-attention architecture, replacing the recurrent layers. 
SAKT \cite{pandey2019selfattentive} model identifies the concepts learnt from past history relevant to the current concept to predict the mastery. The data sparsity problem is handled well by the SAKT model as it predicts based on relatively fewer activities from the past and identifies the relevance between the knowledge concepts.

The prediction of the student's performance on completed exercises are evaluated by assigning attention weights to each of them and visualising which relevant past activities are leveraged by the network to solve the current exercise. In every self-attention layer of SAKT, each query is an exercise embedding vector, and key and values are interaction embedding vectors. The different layers of SAKT includes the embedding layer, multi-head attention and feed forward layer. The exercises are embedded into a collection of attention networks concatenated and assigned weights used for prediction.

\subsection{Experimental setup}
\label{sec:expt}

In this section, we present the results from our study on the subset of the ASSISTMENT 2009 data set described earlier.

The whole dataset is grouped into a user dependent data with the student doing the problems(user\_id), information of the questions he/she has answered(problem\_id), the school where the problem was assigned(school\_id) and the correctness of the answers (correct) taken as input for the train data table. Every student's history of activities is represented as a sequence of one row per student having the same length. The model takes each sequence of questions as an input. For sequence lengths shorter than the given length, the data is appended with zeros and for exceeding sequence lengths, they are truncated to the sequence\_length defined in the model and split as another sample with the remaining sequence length.

\begin{figure}[h]
\centering
\includegraphics[width=0.4\textwidth]{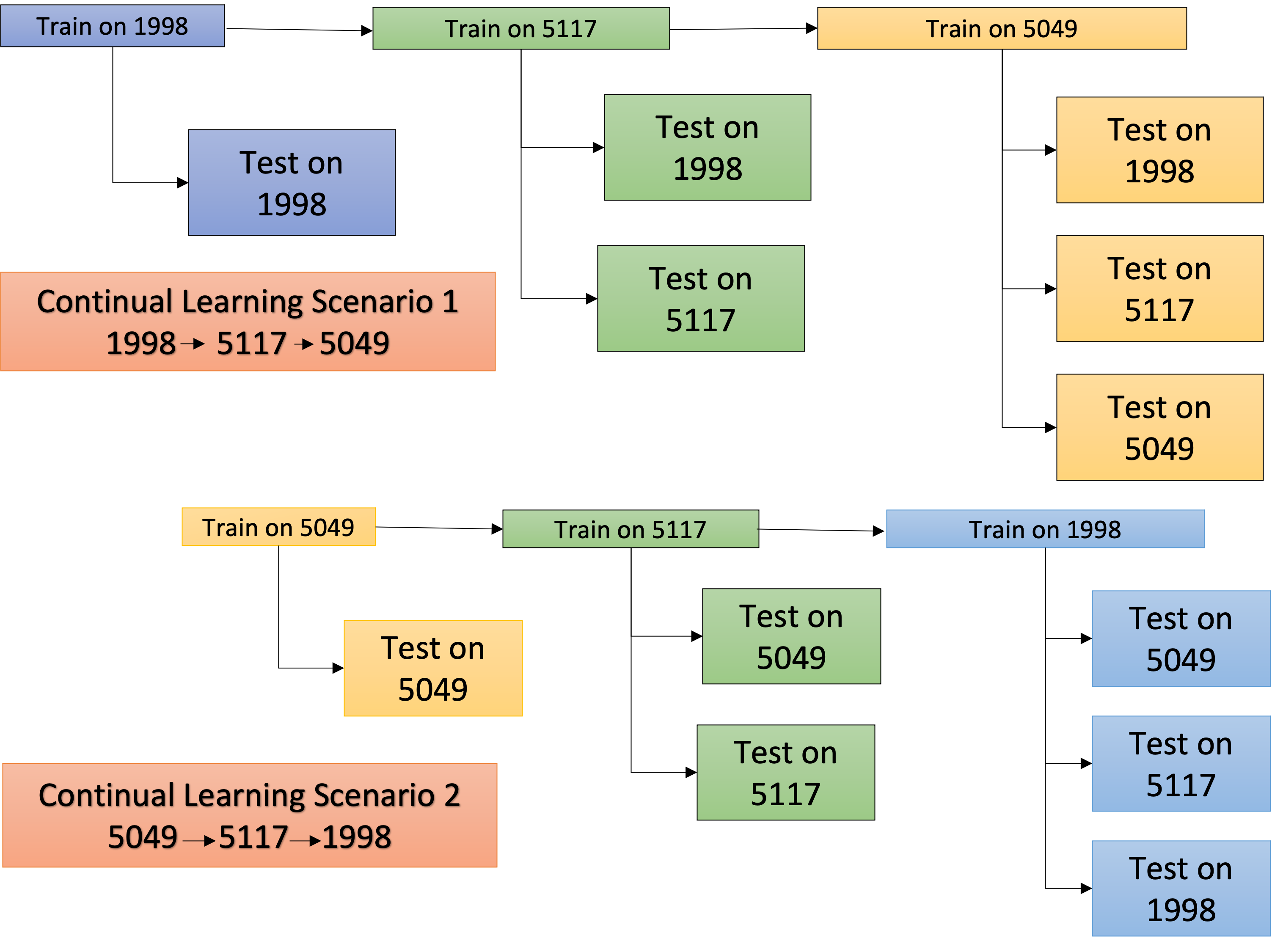}
\caption{Flowchart depicting the process of Continual learning scenario on 1998-5117-5049 and 5049-5117-1998}
\label{ContL}
\end{figure}

The predicted accuracy (ACC), area under the curve (AUROC) and precision-recall curve (AUPRC) values are obtained individually for each current task and learned continually from the previous tasks. The training is done sequentially on the 3 tasks and the testing on the current and the previous tasks, as shown in \textit{Figure \ref{ContL}}. 
The two scenario's considered for continual learning are:
\begin{itemize}
\item Sequence of Scenario 1: $1998\rightarrow5117\rightarrow5049$
\item Sequence of Scenario 2: $5049\rightarrow5117\rightarrow1998$

\end{itemize}
In Scenario 1, the model is trained on task1(1998) and tested on 1998 itself. After continually training on task2(5117), the model is tested on task1(1998) and task2(5117). Finally, after training on task3(5049), the model is tested on all the three tasks(1998\&5117\&5049) as depicted in Figure \ref{ContL}. 
The performance of the binary classification setting of the prediction task is compared by obtaining the AUROC metric. As the AUROC metric does not reflect on the ability of the model to predict the minority class, we also report the AUPRC. 

\subsection{Model Details}
The model parameters are trained using the Binary Cross-Entropy Loss, with a 5-fold cross validation to split the train and test set. An embedding size of 128 with 2 encoder and decoder layers are used. SGD optimizer with a momentum of 0.99 and learning rate of 0.001 initially, scheduled to have a maximum warm up stage of 50 steps upto 0.002 is used. The maximum sequence length was set to 30 in the original SAKT paper and we have followed the same max\_seq\_len = 30.

\section{Experimental Results}
\label{sec:results}

Before the results for the two proposed scenarios are presented, the results for disjoint and joint training is given here. In disjoint training, the model is trained on data from one school (task) and evaluated on all the tasks of the other schools (tasks). In joint training, the model is trained with the data from all schools (tasks) and evaluated on all of them. Hence, the joint training results can be taken as the upper bound.

\begin{table}[htp!]
\centering
\includegraphics[width=0.51\textwidth]{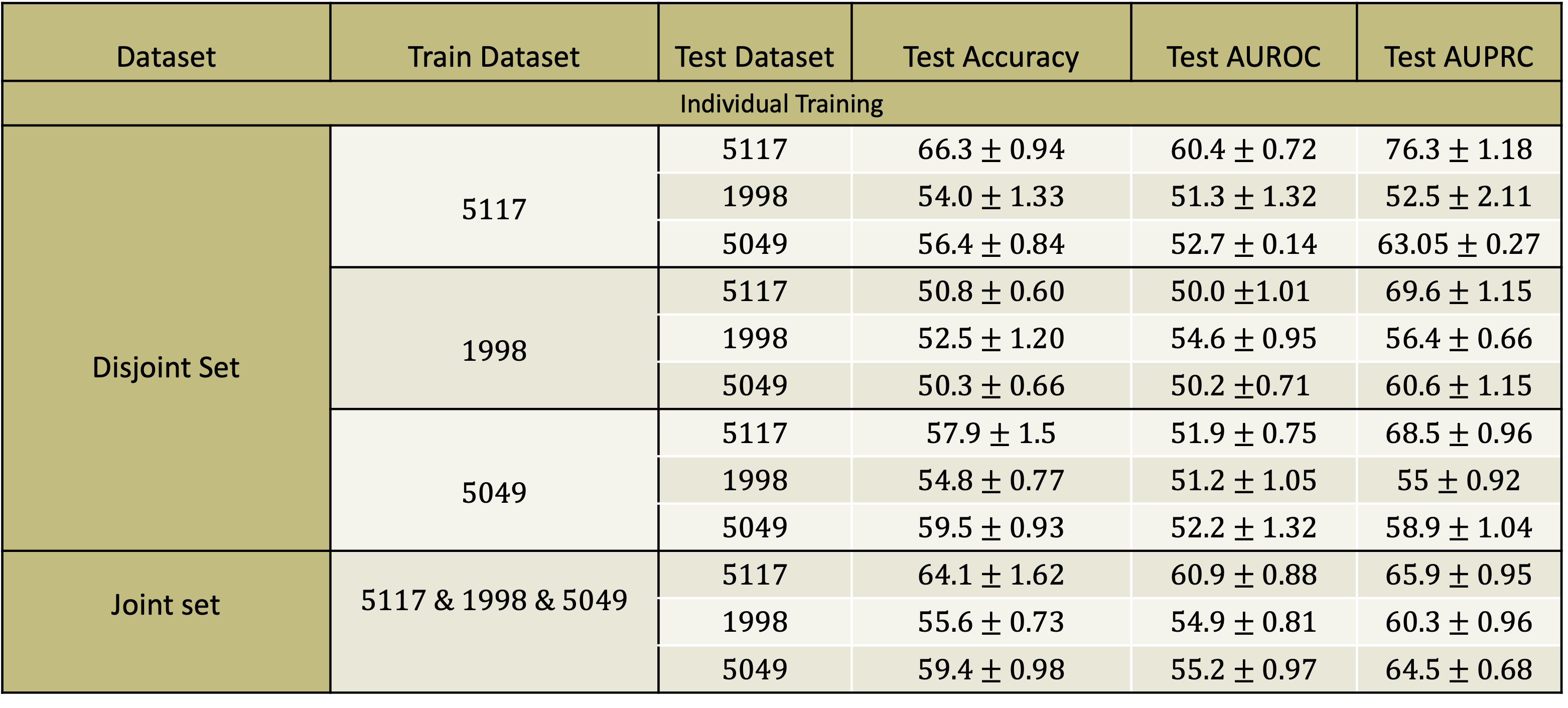}
\caption{Results of Individual training of the 3 tasks and the joint training of all the three tasks}
\label{tab:disjoint_jt}
\end{table}

Table \ref{tab:disjoint_jt} shows the results of the disjoint and joint training of the proposed problem presented earlier. The AUROC/AUPRC for task 5117 is lower (AUROC: $\approx50\%$, AUPRC: $\approx70\%$) when trained on 1998 as compared to being trained on 5117 (AUROC: $\approx60\%$, AUPRC: $\approx76\%$). Similarly, the AUROC/AUPRC for task 1998 is lower (AUROC: $\approx51\%$, AUPRC: $\approx53\%$) when trained on 5117 as compared to being trained on 1998 (AUROC: $\approx55\%$, AUPRC: $\approx56\%$). These prediction results reveals the data drift between 5117 and 1998 seen in Figure \ref{tsne}. The AUROC in the joint training is higher or similar than those in the disjoint training. This indicates that the data from multiple schools are complementary and helps to improve in the overall performance in the tasks.
\begin{figure}[htp!]
\centering
\includegraphics[width=0.20\textwidth]{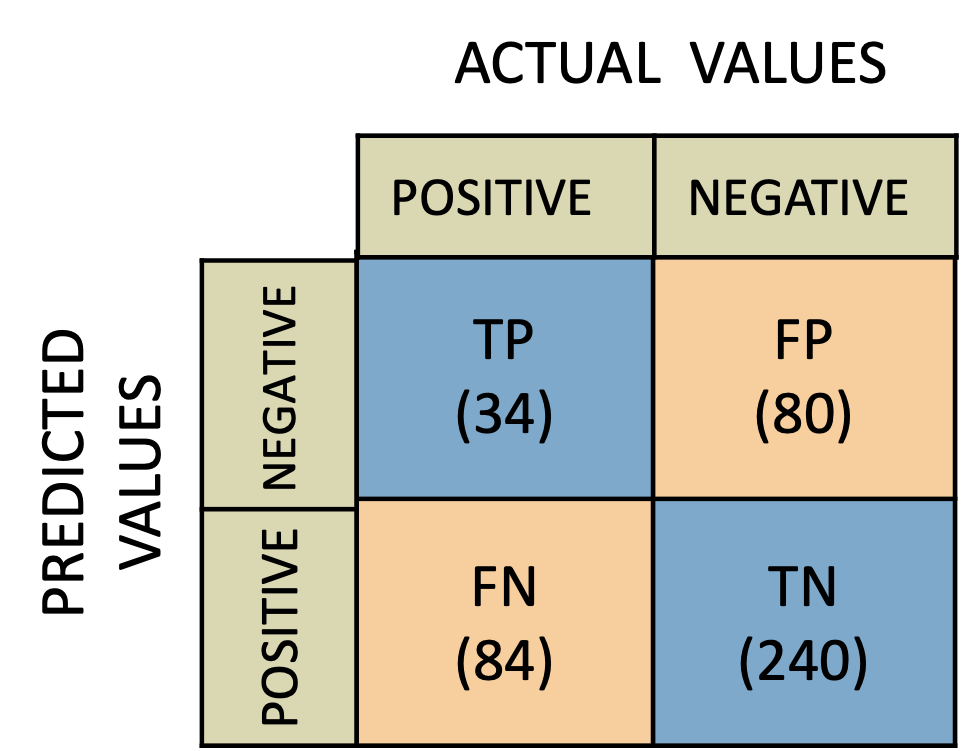}
\caption{Confusion Matrix}
\label{cm}
\end{figure}

\begin{figure*}[h!]
\centering
\centerline{\includegraphics[width=16.5cm]{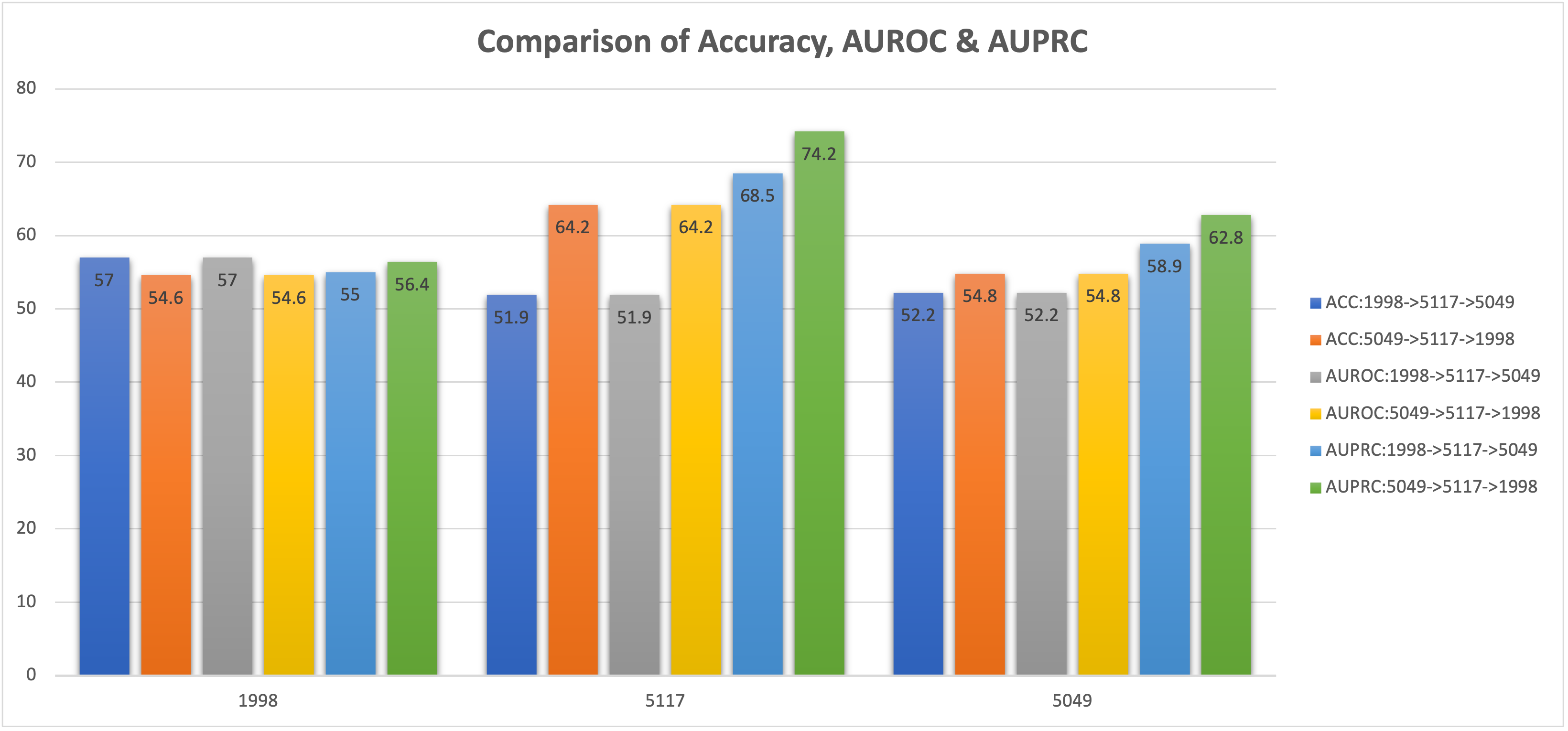}}
\caption{Illustration of Accuracy, AUROC and AUPRC values for the 3 tasks 1998,5117,5049 for the two CL scenarios mentioned as 1998-5117-5049 and 5049-5117-1998. }\medskip
\label{bar}
\end{figure*}

The ACC, AUROC and AUPRC values calculated for continual learning on the two scenarios is shown in the below Figure \ref{bar}. It can be observed that the ACC and AUROC values of CL scenario 1 is better for 1998 when comparing with CL scenario 2. Whereas, the performance of the other two tasks of 5117 and 5049, is better in scenario 2 when compared to scenario 1. 
We can observe that the accuracy of the tasks is affected by the order of the sequence resulting in lesser accuracy of task 1998 in scenario 2 than scenario 1. This might be because for the scenario 2, 1998 task (3rd task) is continually learning from the 5049 and 5117 tasks whereas the learning starts from 1998 (1st task) as in the case of scenario 1.

The imbalance in data is reflected by higher AUPRC values than AUROC as the number of samples in the true positive class is much lesser than the samples of the true negative class. This can be observed from the confusion matrix illustrated in Figure \ref{cm}. 

The performance of scenario 2 is better in task2 (5117) compared scenario 1 
as 5117 uses the shared knowledge learnt in 5049 which helps in better prediction. The might be because 5049 (learnt task in scenario 2) has a much greater number of responses (19106) as compared to 1998 (learnt task in scenario 1), where the number of responses is only 5617 (refer to Table \ref{tab:data_char}). The greater amount of data could help the model to generalize better to the characteristics of the knowledge tracing problem. 

\begin{figure}[h!]
\centerline{\includegraphics[width=9.0cm]{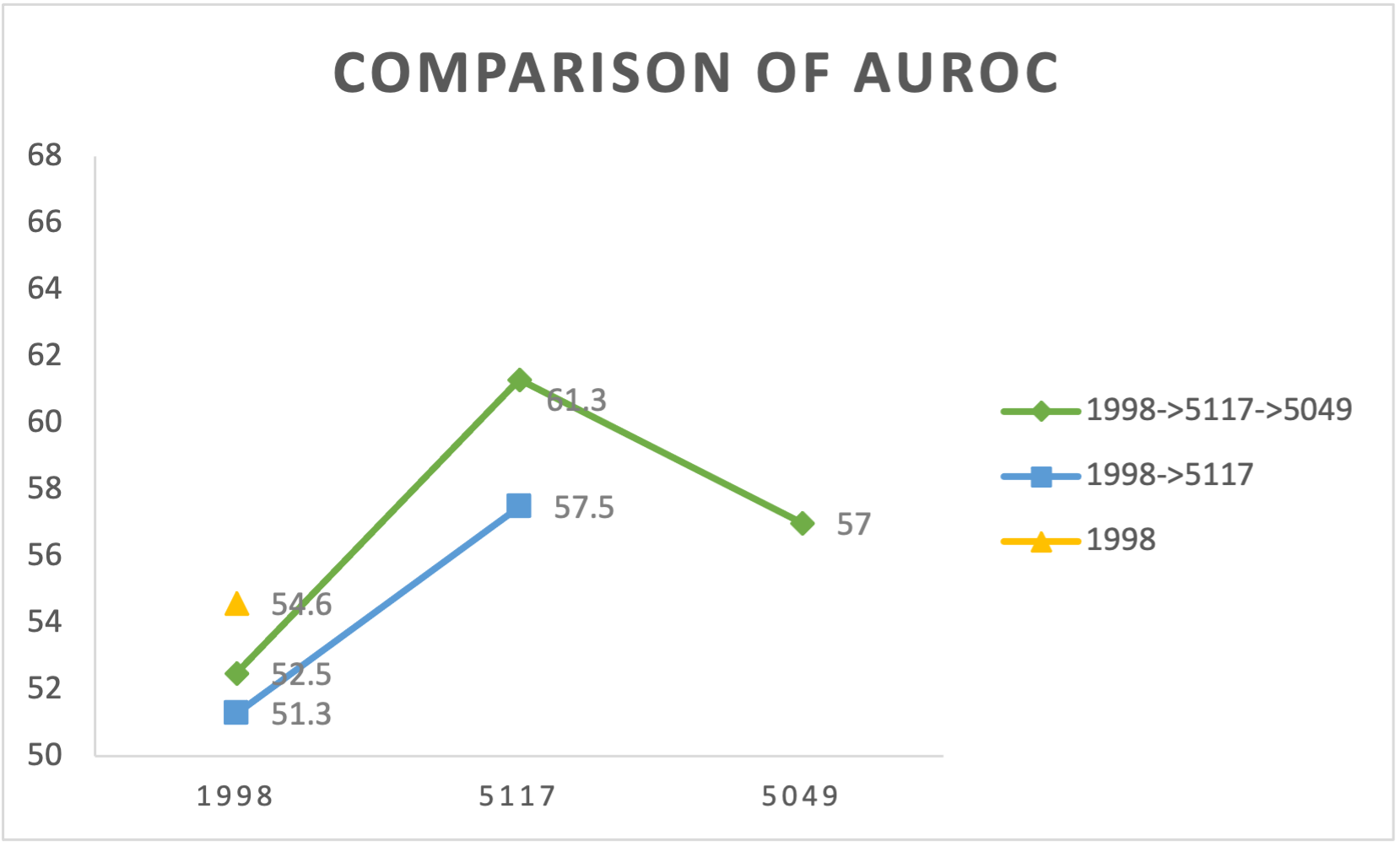}}
\caption{Comparison of AUROC values for the continual learning scenario 1}\medskip
\label{AUROC1}
\end{figure}
\begin{figure}[h!]
\centerline{\includegraphics[width=9.0cm]{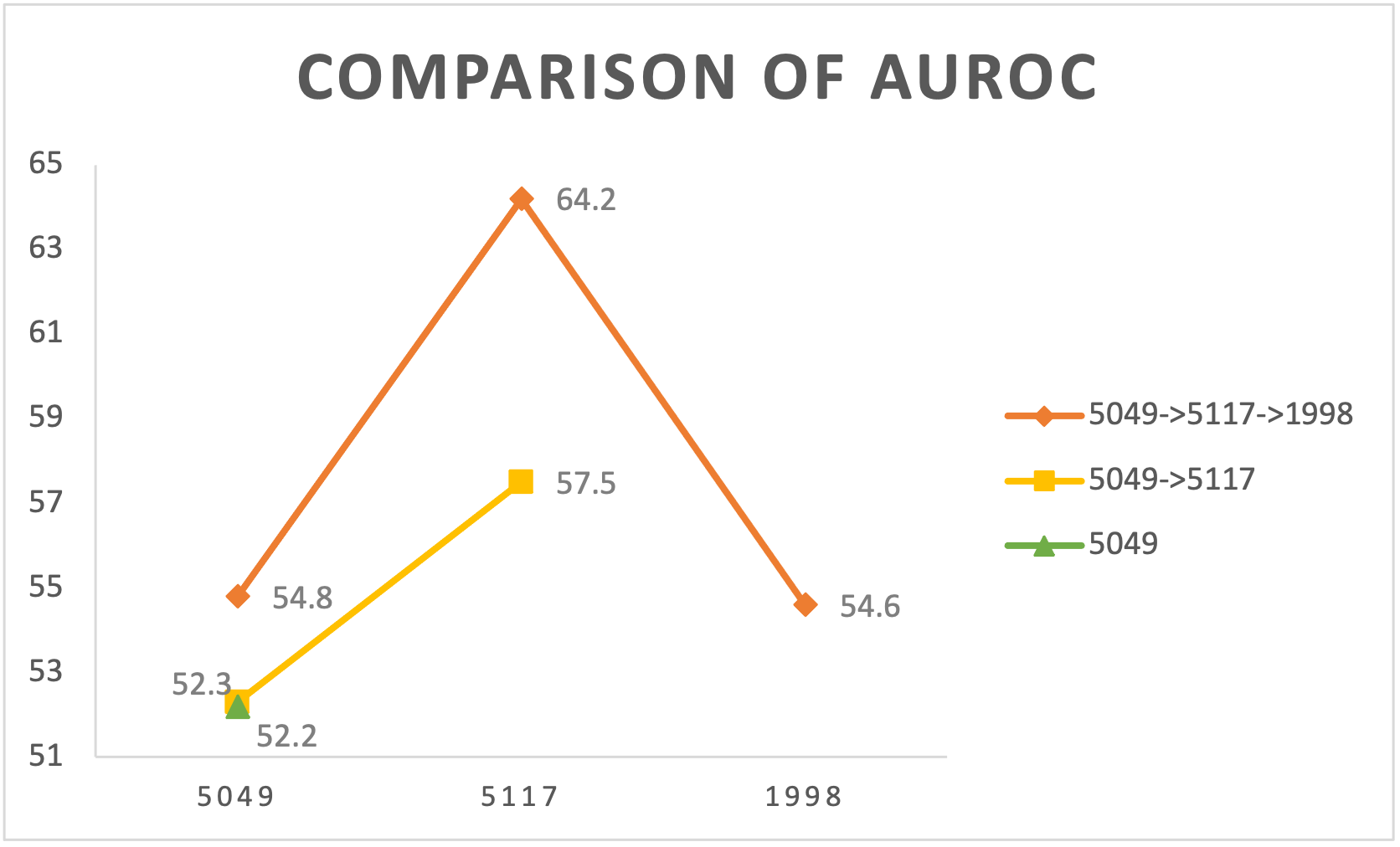}}
\caption{Comparison of AUROC values for the continual learning scenario 2}\medskip
\label{AUROC2}
\end{figure}

The line plot in Figure \ref{AUROC1} illustrates the CL scenario 1 for the three tasks. The yellow triangle(AUROC of 54.6) represents training and testing on task 1, the blues squared line is the result of after training on 5117 and testing on both 1998 and 5117. The green diamond line is after training on 5049 and tested on all the 3 tasks. The AUROC value decreases from the first task to the second but increases for the third. This indicates that the model have forgotten some of the previously learnt knowledge in 1998 whilst learning the current task of 5117. However, the learning of the task 5049 helps in improving the AUROC value of 1998 after the training of 5049 because of the similarity of problem\_ids between tasks 5049 and 1998 (reflected in Figure \ref{tsne}).

The second scenario of CL is illustrated in Figure \ref{AUROC2} which is training and testing on task 3(5049) represented by the green triangle, then continually learning on 5117 (yellow squared line) and then 1998 (Red diamond line). The AUROC trend is increasing in the order of training the tasks, i.e., when the 3rd task (1998) is trained, and tested on all the three tasks, the AUROC is much higher than the previous trained AUROC which implies that the learning from the current task helps in improving the prediction performance in the previous tasks.

Comparing the two scenarios, the scenario 2 seems to edge out in performance when compared to scenario 1. As the continual learning of the tasks depends on the distribution of data, starting training with a bigger dataset results in a much stable model then using a smaller dataset as the first task. 
5049 is a large dataset with more data at the start leading to a much generalizable and stable model rather than starting with lesser data like 1998 that will probably converge to a local optima than a global one. However, one may not be able to always start training on a large dataset due to data availability. In the possible case of knowledge learnt being forgotten (from 1998-5117 in scenario 1), one can implement continual learning strategies such as regularisation, which can be the future work of exploration.

\subsection{Ablation Study}
From the experiments and results of the continual learning of scenario 1 \& 2, it can be observed that there is little overlap between the tasks 1998 and 5117. Here, we explore the data drift in 1998-5049 and 5117-5049.

\begin{table}[htp!]
\centering
\includegraphics[width=0.51\textwidth]{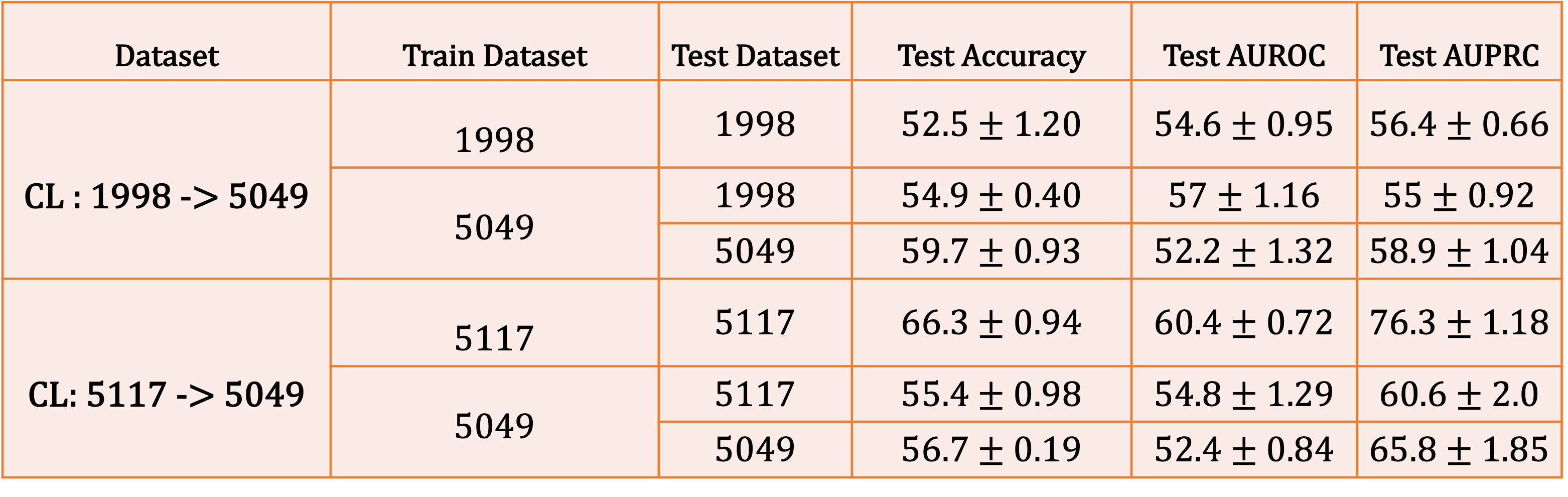}
\caption{Results of ablation study on tasks 1998-5049 and 5117-5049}
\label{tab:ab_tab}
\end{table}
The ablation study is done between tasks 1998-5049 and 5117-5049 as illustrated in Table \ref{tab:ab_tab} in a similar way explained above to see how shared representations affect the continual learning strategy. Comparing the AUROC metric for both the scenarios 3 and 4 (1998-5049 and 5117-5049), the AUROC of task1 1998 increases when continually trained on 5049 but the AUROC of 5117 decreases when continually trained on 5049. The overlapping between the tasks is the major contributing factor for this observation. This supports the remark made in the above sections that there is less similarity in the data between the tasks 5117 and 5049 and that 1998 is more similar to 5049.

\section{Conclusion}
\label{sec:concl}
This paper provides a novel study on incremental knowledge tracing of multiple schools while preserving the privacy of data in each school, through implementation of the the self attentive knowledge tracing model.
We introduce the continual learning framework which helps the model to learn continually without forgetting the knowledge gained in the previous tasks. This performance of the incremental knowledge tracing with SAKT is demonstrated on the ASSISTment 2009 dataset. The effectiveness of our approach is depicted by the similar performances found when all data from multiple schools are made available. To further improve our performance, we can explore the incorporation of regularization strategies to avoid catastrophic forgetting. 


\bibliography{aaai22}

\end{document}